\documentclass[useAMS, usenatbib]{mn2e} 

\usepackage{graphicx}

\title{From cosmic ray source to the Galactic pool}
\author[K.M. Schure \& A.R. Bell]{K.M. Schure$^{1}$\thanks{E-mail: kschure@gmail.com} and A.R. Bell$^{1}$\\
$^{1}$Department of Physics, University of Oxford, Clarendon Laboratory, Parks Road, Oxford OX1 3PU, United Kingdom}

\begin{document}

\newcommand\araa{{ARA\&A}}
\newcommand\apj{{ApJ}}
\newcommand{\apjl}{ApJL}
\newcommand\apjs{{ApJS}}
\newcommand\aap{{A\&A}}
\newcommand\mnras{{MNRAS}}
\newcommand\rmxaa{{Rev. Mexicana Astron. Astrofis.}}
\newcommand\nat{{Nature}}
\newcommand\physrep{{Phys.~Rep.}}
\newcommand\memsai{{MmSAI}}
\newcommand\ssr{{Space Science Rev.}}

\date{\ldots; \ldots}
\pagerange{\pageref{firstpage}--\pageref{lastpage}}\pubyear{2013}
\maketitle
\label{firstpage}

\begin{abstract}
The Galactic cosmic ray spectrum is a remarkably straight power law. Our current understanding is that the dominant sources that accelerate cosmic rays up to the knee ($3 \times 10^{15}$~eV) or perhaps even the ankle ($3 \times 10^{18}$~eV), are young Galactic supernova remnants. In theory, however, there are various reasons why the spectrum may be different for different sources, and may not even be a power law if nonlinear shock acceleration applies during the most efficient stages of acceleration. We show how the spectrum at the accelerator translates to the spectrum that make up the escaping cosmic rays that replenish the Galactic pool of cosmic rays. We assume that cosmic ray confinement, and thus escape, is linked to the level of magnetic field amplification, and that the magnetic field is amplified by streaming cosmic rays according to the non-resonant hybrid or resonant instability.
When a fixed fraction of the energy is transferred to cosmic rays, it turns out that a source spectrum that is flatter than $E^{-2}$ will result in a $E^{-2}$ escape spectrum, whereas a steeper source spectrum will result in an escape spectrum with equal steepening. This alleviates some of the concern that may arise from expected flat or concave cosmic ray spectra associated with nonlinear shock modification.
\end{abstract}


\section{Introduction}
\label{sec:intro}

Consensus is building that supernova remnants (SNRs) are sources of cosmic rays, and most likely constitute the dominant source for the component of cosmic rays (CRs) that is believed to originate within our own Galaxy. Magnetic field amplification needs to occur to aid the acceleration to such high energies, and in general is considered important in scattering the cosmic rays in the vicinity of the shock in order to enable the many shock crossings needed for acceleration to highly relativistic speeds. The efficiency that is needed is thought to go hand-in-hand with non-linear modification of the shock structure, resulting in a concave spectrum, and a rather flat spectrum at the highest energies. The fact that the CR spectrum as measured on Earth is rather steep and shows no such concavity or flattening has been considered troublesome. In this paper we will show how this does not necessarily need to be contradictory: we find that the escape spectrum of CRs only mimics the source spectrum when the spectrum is steeper than the $E^{-2}$ power law. When the source spectrum is flatter than $E^{-2}$, the escape spectrum is still as steep as $E^{-2}$.

Recently, there has been increasing interest in the translation between source spectrum, being the spectrum produced at the shock, and the escape spectrum \citep{2011Drury, 2010Capriolietal,2010Ohiraetal}. Early discussions on this topics used to focus on explaining that the Galactic cosmic ray spectrum is a nearly straight power law. However, recent data showing more detail and differences between hydrogen, helium, and other species, confirm that the picture is more complicated \citep{2013Ptuskinetal}. Also, gamma ray observations, and the connection with molecular clouds, have increased awareness that part of the cosmic rays that we detect in the source may in fact already be the ones escaping upstream. A statistical model of the expected gamma ray sources assuming a universal source spectrum favours a spectrum steeper than $E^{-2.1}$ \citep{2013Cristofarietal}. All of these developments require a more thorough understanding of the changes the spectrum may undergo from source to detection. Major improvement on the high energy end of the gamma ray spectrum of SNRs may be achieved once CTA will be online \citep{2013CTA}.

Despite the lack of evidence from observations, from the theoretical standpoint there are compelling reasons why a concave spectrum, and thus a flat spectrum at the high-energy end is to be expected. When CR acceleration is efficient and the cosmic ray pressure at the shock is significant, this is expected to alter the structure of the shock. As a result, the low-energy cosmic rays will only probe part of the compression at the shock, while the high-energy particles `see' the full shock compression, which in this case may also be higher than a factor 4. This results in a concave spectrum \citep{1979Eichler,1980Blandford,1981DruryVoelk,1987Bell,1984EllisonEichler,2009Kangetal,2009Vladimirovetal,2010Capriolietal}, with a spectral index that varies between steeper and shallower than the canonical $E^{-2}$ power law spectrum. In fact, when SNR should be responsible for replenishing the Galactic cosmic ray population, it is expected that, at least for the most efficient stages and sources in acceleration, such nonlinear shock modification must occur. 

There are other reasons why the CR source spectrum at SNR shocks may deviate from the canonical $E^{-2}$ power law. A higher (lower) compression ratio at the shock than the typical factor four will result in flattening (steepening). There may be various causes why compression ratios may differ, including radiative effects and low-Mach number shocks. In addition, oblique magnetic fields can alter the spectral power law index \citep{2011Belletal}.

A number of authors have shown that the dynamics of the shock essentially determines the escape spectrum, and in the Sedov phase this naturally leads to a $E^{-2}$ power law \citep[e.g.][]{2011Drury, 2010Capriolietal, 2010Ohiraetal,2013Belletal}. \citet{2010Ptuskinetal} take into account the change in maximum energy to derive a total produced spectrum and convolve it with the expected energy-dependent escape in the Galaxy to arrive at the Galactic CR spectrum, which then looks rather like the source spectrum, but steeper over the complete energy range. In this paper, we will highlight the importance of the dynamics, but furthermore we will demonstrate why the spectrum that escapes from SNRs is always $E^{-2}$ or steeper, and never consistently flatter, provided the energy transferred to CRs at the shock is a fixed fraction of the kinetic energy.

\section{Escape spectrum}

In this paper we build on previous work \citep[][hereafter SB13]{2013SchureBell}, which looked into the maximum attainable cosmic ray energy, on the premise that the magnetic field needs to be amplified to levels that confine the maximum-energy cosmic rays. The escaping component of cosmic rays trigger a return current that amplifies the magnetic field, and feeds into the Galactic cosmic ray spectrum. While previously we only considered a source spectrum that had the typical $E^{-2}$ power law, in this paper we investigate the effects of different source spectra on the Galactic cosmic ray spectrum.

While we will not dwell on the reason for spectral steepening or flattening at the source, we will investigate the result of a different source spectrum on the expected Galactic cosmic ray spectrum. In order to isolate the argument we will limit ourselves to consider the following:
We will look at young supernova remnants ($t<5000$~years), and assume that the maximum CR energy is determined by their confinement resulting from self-generated magnetic field growth \citep{2004Bell}. The magnetic field growth can be the result of the non-resonant hybrid (NRH) instability or the resonant instability \citep[for an overview see][]{2012Schureetal}, and we assume that a fixed number of e-folding times, represented by the growth rate $\gamma$ multiplied with the time $\tau$, is necessary for confinement.

The definition of `escape spectrum' is not always clear, a nice discussion of which has recently been presented by \citet[][]{2011Drury}. What we mean here by the escape spectrum is that part of the cosmic ray population that propagates upstream and drives the plasma instabilities that confine the `non-escaping' CR. 

Given the assumption that cosmic ray escape is determined by magnetic field amplification by $\exp(\gamma \tau)$, we can determine the cumulative escape spectrum over the age of the remnant (SB13). We assume that at any given time, the escaping particles constitute a delta function in energy. The energy is determined by two parameters: the number of cosmic rays needed to escape in order to generate a strong enough current to amplify the magnetic field, and by the energy available in the cosmic ray spectrum at the shock itself. The latter we parameterize by assuming that a fixed fraction $\chi$ of the shock kinetic energy density is transferred to cosmic rays, such that $U_{\rm cr}=\chi \rho u_s^2$. In the case where $30$\% of the downstream thermal energy is converted to cosmic rays, $\chi=0.34$. The number of cosmic rays now depends on the spectral distribution of cosmic rays at the shock. 

In SB13 we only considered a $E^{-2}$ power law source spectrum. Here we will reiterate this analysis and extend the derivation to allow for deviations from a $E^{-2}$ source spectrum. We will keep the slope fixed over energies, i.e. we neglect concavity or convexity of the spectrum at the shock.

In the case of the canonical $E^{-2}$ power law, the number density of cosmic rays above an energy $E$ is given by:
\begin{eqnarray}
\frac{N}{N_0}=\left( \frac{E}{E_0} \right)^{-1},
\end{eqnarray}
where $N_0$ is the number density of cosmic rays above a reference energy $E_0$.

The current at the shock as a function of the shock velocity $u_s$ and shock radius $R_s$, such that the magnetic field is amplified by $\gamma \tau$ e-foldings, for a maximum growth rate $\gamma=j/c\sqrt{\pi/\rho}$ \citep{2004Bell}, needs to be (see Eq.~5 in SB13): 
\begin{eqnarray}
\label{eq:j}
j(R_s)=\frac{s_0 \gamma \tau c u_s}{R_s}\sqrt{\frac{\rho}{\pi}}.
\end{eqnarray}
Here $\rho$ is the density of the unperturbed medium directly upstream of the shock, and the factor $s_0$ is $1$ in the case of a circumstellar medium (CSM) that is shaped by a stellar wind, where the density drops with the radius as $\rho \propto R^2$, and $s_0=2$ in the case of a homogeneous interstellar medium (ISM). The reason for this is the dependency of the growth rate for magnetic field amplification on the density, which is higher for a lower density, thus making amplification more efficient for a CSM environment. 

We see from Eq.~\ref{eq:j} that the important factor that determines the confinement of cosmic rays is the current that results from escaping cosmic rays. This can be made up by a low CR number density streaming at a high drift velocity, or a higher number density with a lower drift velocity. To assess what that means for the equivalent number density $N_{\rm esc}$ of cosmic rays that needs to escape at the shock, we find that:
\begin{eqnarray}
N_{\rm esc}=\frac{j}{q \zeta u_s},
\end{eqnarray}
where $\zeta u_s$ is the drift velocity of escaping CRs as they leave the shock, and $q$ is the unit charge. This number density has to be equal to the number density of cosmic rays as can be calculated from the spectrum. This connects the shock evolution to a representative number density at the shock:
\begin{eqnarray}
\label{eq:nesc_1}
N_{\rm esc} =\frac{s_0 \gamma \tau c}{\zeta q R_s}\sqrt{\frac{\rho}{\pi}}.
\end{eqnarray}

We can furthermore use the constraint of the energy that is available in cosmic rays to find the maximum energy:
\begin{eqnarray}
N(E) dE &=& -\frac{N_0}{E_0} \left( \frac{E}{E_0}\right)^{-2} dE\\\nonumber
U_{cr}&=&\int^{E_{\rm max}}_{E_0} N(E)E dE=-\frac{N_0}{E_0} \left( \frac{E}{E_0}\right)^{-2} E dE\\\nonumber
& =& -N_0 E_0 \ln\left(\frac{E_{\rm max}}{E_0}\right) \\\nonumber
&=& \chi \rho u_s^2.
\end{eqnarray}
Alternatively, we can express this in terms of the energy flux of the cosmic rays that escape from the system with a streaming velocity $\zeta u_s$:
\begin{eqnarray}
Q_{\rm esc}=\chi \zeta \rho u_s^3.
\end{eqnarray} 

From this we can deduce the normalisation factor $N_0 E_0$ to be
\begin{eqnarray}
N_0 E_0&=&-\frac{\chi \rho u_s^2}{\ln(E_{\rm max}/E_0)}.
\end{eqnarray}
Using this in Eq.~\ref{eq:nesc_1} we find that:

\begin{eqnarray}
N_{\rm esc}=\frac{N_0 E_0}{E_{\rm max}} = -\frac{\chi \rho u_s^2}{E_{\rm max}\ln(E_{\rm max}/E_0)}=\frac{s_0 \gamma \tau c}{\zeta q R_{s}}\sqrt{\frac{\rho}{\pi}},
\end{eqnarray}
from which follows the maximum energy as a function of the evolution of the shock wave:
\begin{eqnarray}
E_{\rm max}=\frac{u_s^2 R_s q \sqrt{\rho \pi}}{s_0 c \gamma \tau \ln(E_{\rm max}/E_0)}\left(\frac{Q_{\rm esc}}{\rho u_s^3}\right).\label{eq:Ebeq0}
\end{eqnarray}
This is similar to the equation used in SB13 apart from the fraction $\zeta$.  Note again that it is the energy flux $Q_{\rm esc}$ relative to $\rho u_s^3$, rather than the energy density $U_{\rm cr}$, that determines $E_{\rm max}$.

We now extend the analysis to a spectrum that is steepened by a certain amount $\beta$, such that for a spectrum steeper than $E^{-2}$, $\beta > 0$, and a flatter spectrum is obtained for $\beta<0$. This analysis is valid so long as the spectrum doesn't get flatter than $-1$, or $\beta>-1$.

The spectrum and the energy contained in cosmic rays are now given by:
\begin{eqnarray}
\frac{N}{N_0}=\left( \frac{E}{E_0} \right)^{-(1+\beta)}
\end{eqnarray}

\begin{eqnarray}
N(E) dE &=& -\frac{N_0}{E_0} \left( \frac{E}{E_0}\right)^{-(2+\beta)} dE\\\nonumber
U_{\rm cr}&=&\int^{E_{\rm max}}_{E_0} N(E)E dE=-\frac{N_0}{E_0} \left( \frac{E}{E_0}\right)^{-(2+\beta)} E dE\\
& =& \frac{N_0 E_0}{\beta} \left(\left(\frac{E_{\rm max}}{E_0}\right)^{-\beta}-1\right),\label{eq:ucrbeta}
\end{eqnarray}
from which we can deduce that the normalisation factor, now $N_0 E_0^{1+\beta}$, is:
\begin{eqnarray}
N_0 E_0^{1+\beta}&=&\frac{\beta \chi \rho u_s^2}{E_{\rm max}^{-\beta}-E_0^{-\beta}}.
\end{eqnarray}
This can be used in Eq.~\ref{eq:ucrbeta} to normalise the number density, such that:

\begin{eqnarray}
\label{eq:nescbeta}
N_{\rm esc}=\frac{N_0 E_0^{1+\beta}}{E_{\rm max}^{1+\beta}} =\frac{\beta \chi \rho u_s^2}{E_{\rm max}^{1+\beta}(E_{\rm max}^{-\beta}-E_0^{-\beta})}.
\end{eqnarray}
For young SNRs $E_{\rm max} \gg E_0$, which means that the relative contributions of $E_{\rm max}^{-\beta}$ versus $E_0^{-\beta}$ in the denominator, and therefore the slope of the escape spectrum, depends strongly on whether $\beta$ is positive or negative. Interestingly, while a steeper source spectrum causes the escape spectrum to be steeper, a flatter source spectrum does {\em not} result in a flatter escape spectrum. 

Meanwhile our requirement on the escape current remains unchanged, so we can equate Eq.~\ref{eq:nescbeta} to Eq.~\ref{eq:nesc_1} such that for a nonzero value of $\beta$ we find the escape energy as a function of the shock dynamics for the different power laws:
\begin{eqnarray}
\label{eq:Enonlinear}
E_{\rm max}=\left(\frac{\beta u_s^2 R_s q \sqrt{\rho \pi}}{s_0 c \gamma \tau \left(E_{\rm max}^{-\beta}-E_0^{-\beta}\right)}\left( \frac{Q_{\rm esc}}{\rho u_s^3}\right)\right)^{1/(1+\beta)}.
\end{eqnarray}
In the different regimes, this simplifies to approximately:

\begin{eqnarray}
\beta > 0: \quad&& E_{\rm max}= \left(\frac{\beta u_s^2 R_s q \sqrt{\rho \pi}}{s_0 c \gamma \tau E_0^{-\beta}}\left( \frac{Q_{\rm esc}}{\rho u_s^3}\right)\right)^{1/(1+\beta)}\label{eq:Ebgt0}\\
\beta < 0: \quad&& E_{\rm max}= -\frac{\beta u_s^2 R_s q \sqrt{\rho \pi}}{s_0 c \gamma \tau}\left( \frac{Q_{\rm esc}}{\rho u_s^3}\right)\label{eq:Eblt0}.
\end{eqnarray}

\section{Spectra for a range of SNRs and spectral indices}

In this section we will illustrate the above calculation by evaluating the resulting escape spectrum for different values for $\beta$. In SB13 we elaborated on the developments and growth of the NRH instability and its consequences for the maximum CR energy. In this section we will similarly base our analysis on the result of 1D hydrosimulations performed with the AMRVAC code \citep{2003Keppensetal,2007HolstKeppens} for the evolutionary characteristics of the shock wave. 

We initialize the grid with either a homogeneous interstellar medium (ISM) with a density of $n=0.85$~cm$^{-3}$, such as could be more applicable to Type Ia supernovae (SNe), or with a circumstellar medium (CSM) formed by a stellar wind with a mass loss rate of $10^{-5}$~M$_\odot$~yr$^{-1}$ and wind velocity of $v=4.7$~km~s$^{-1}$, such as may occur during the red supergiant (RSG) phase of a massive star and appropriate for Type II SNe. The total numerical grid extends to a radius of $\sim 13$~pc with a resolution of $\Delta r = 6.5\times10^{14}$~cm. We initialize the ejecta in the first $5 \times 10^{15}$~cm using a high density core with a power law envelope following either $\rho \propto r^{-7}$ for the type Ia (ISM), or $\rho \propto r^{-9}$ for the CSM. The explosion energy and mass are set to $10^{51}$~erg and $1.4$~M$_\odot$ for the Type Ia simulation, and to $2 \times 10^{51}$~erg and $2.5$~M$_\odot$ for the Type II explosion. We follow the evolution for $\sim 5000$~yr or until the forward shock has left the grid.

\begin{figure}
  \centering
  \includegraphics[width=0.5\textwidth]{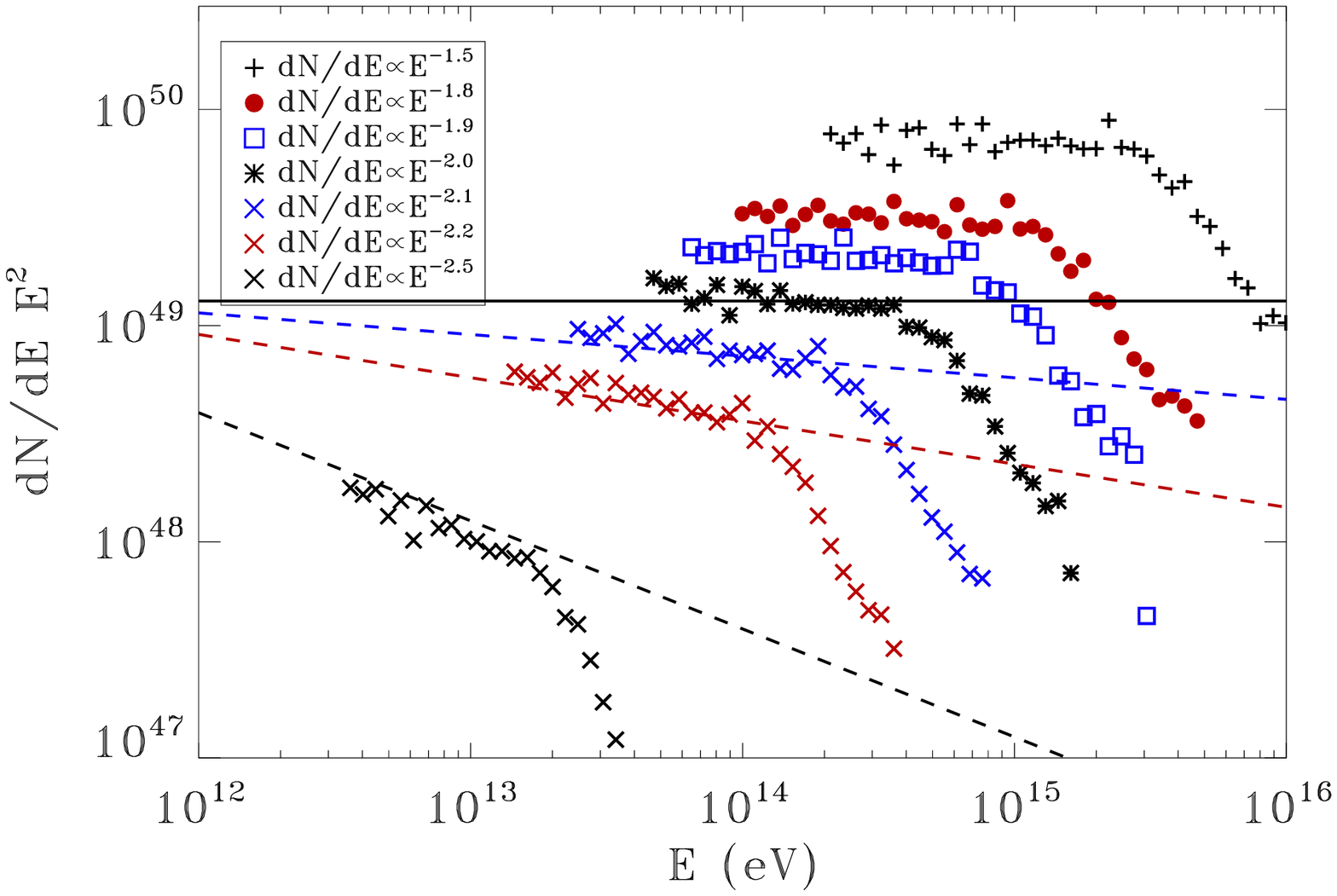}
 \includegraphics[width=0.5\textwidth]{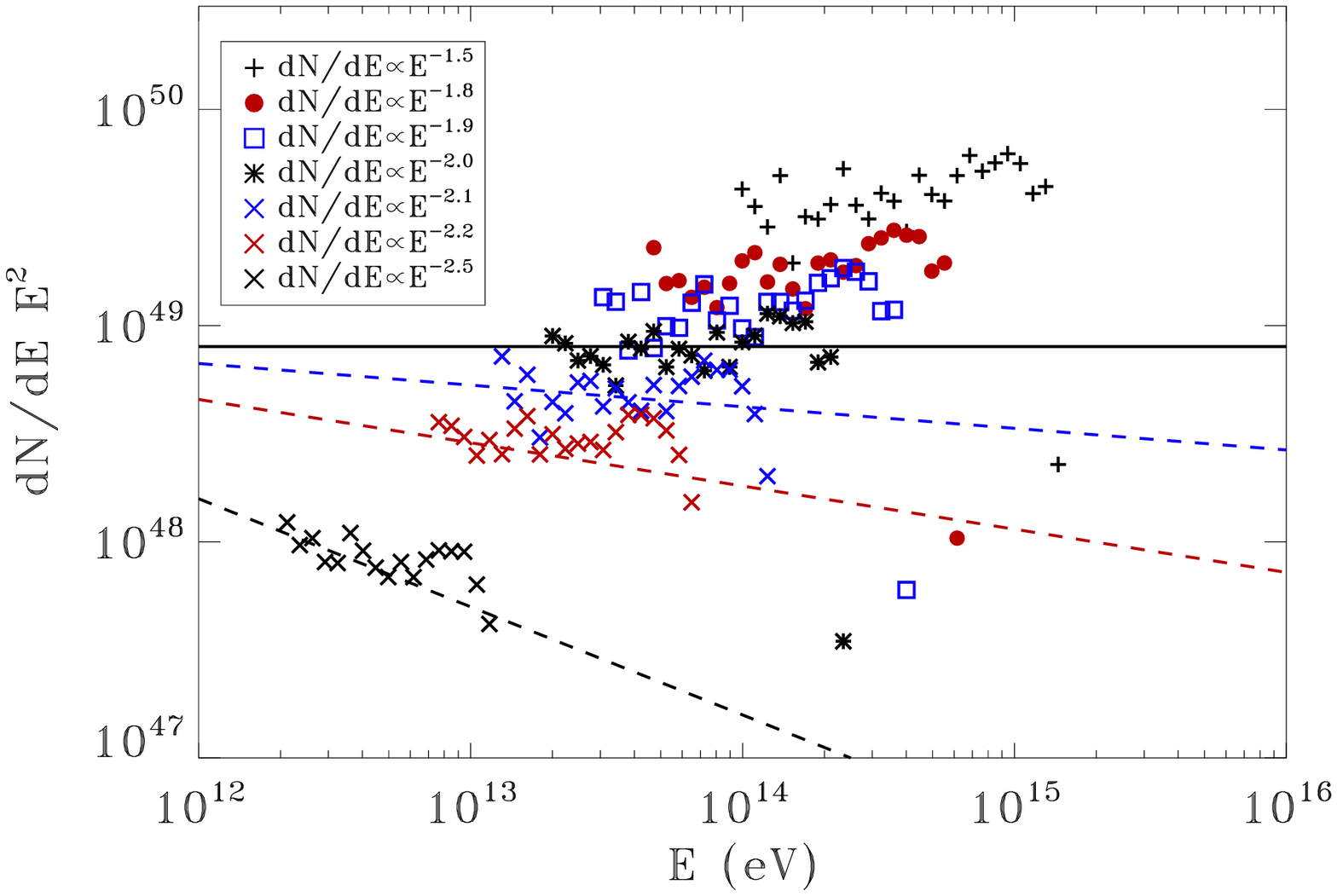}
 \includegraphics[width=0.5\textwidth]{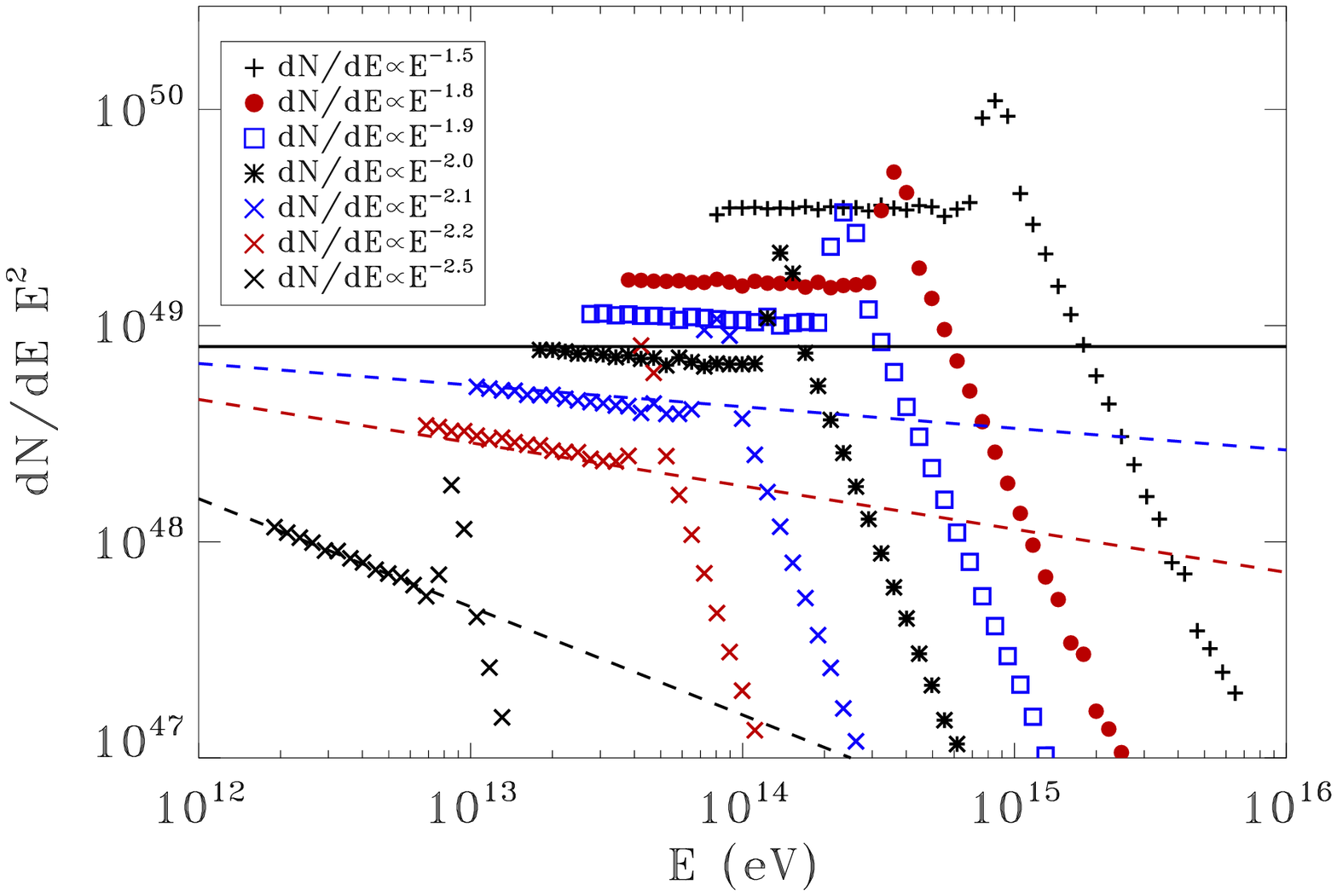}
 \caption[ ] {Spectrum of the escaped cosmic rays for different values of $\beta$, the power law slope of the accelerated particles at the SNR shock. Top: a Type II SN evolving in a CSM shaped by a RSG wind. Middle: a Type Ia SN in a homogeneous medium with $n=0.85$~cm$^{-3}$. Bottom: same SNR as for the middle panel, but with the analytical rather than the numerical solution for the shock evolution. 
    \label{fig:slopes}}
  \end{figure} 

The time evolution of the shock position and velocity as extracted from the simulation results is subsequently used to calculate $E_{\rm max}$ for a number of different values for $\beta$, the slope at the shock, based on the equations described in the previous section. For $\beta > 0$ and $\beta < 0$ we make a first estimate for $E_{\rm max}$ using Eq.~\ref{eq:Ebgt0} and Eq.~\ref{eq:Eblt0}, and use this first estimate to refine the solution by feeding it back into the more precise Eq.~\ref{eq:Enonlinear}, which converges after a few iterations. For $\beta=0$ we set $\ln(E_{\rm max}/E_0)=14$ in our first approximation of the solution in Eq.~\ref{eq:Ebeq0}.

The results are shown in Fig.~\ref{fig:slopes}. We have plotted the escape spectrum multiplied by $E^2$ for the SNR evolution in the CSM (top) and for the ISM (middle panel), for source spectra that follow a power law with indices between $-1.5$ and $-2.5$. Because the spectrum for the ISM scenario shows some fluctuations, possibly because of the difficulty of accurately tracking the shock velocity and the strong dependence of $E_{\rm max}$ thereon, we have also run the analysis using the analytical solution for the shock evolution in an ISM \citep[e.g.][]{1982Chevalier,1999TrueloveMcKee}, the results of which are shown in the lower panel. 

For a source spectrum that is flatter or equal to $E^{-2}$, the escape spectrum follows a $E^{-2}$ power law, while for a steeper spectrum, the escape spectrum is as steep as the source spectrum. This result is most apparent in the solution for the CSM (top panel) and the analytical solution for the ISM (bottom panel), and somewhat less clear in the numerical solution for the ISM (middle panel) because of the numerical noise. The bump at the high-energy end in the ISM models is the result of CR accelerated at the earliest times when $E_{\rm max}$ is relatively low. It takes a longer time for the maximum energy to reach its highest value in a homogeneous ISM (decades, rather than years in the CSM environment). This is accentuated in the analytic case where a large number of CR is injected at early times due to an unrealistically high shock velocity.

\section{Conclusion}
We investigate the connection between the source spectrum and the CR released into the Galaxy. In our analysis we assume that the cosmic rays require magnetic field amplification in the upstream in order to confine them, and a fixed fraction of kinetic energy is transferred to cosmic rays. We find that the escape spectrum does not become flatter than $E^{-2}$, even for a source spectrum flatter than $E^{-2}$. On the other hand, when the spectrum is steeper than $E^{-2}$, the escape spectrum follows the steepness of the source spectrum.
 
This potentially solves one of the problems encountered by non-linear shock acceleration. When cosmic ray acceleration is efficient, the cosmic ray pressure alters the shock structure, causing the cosmic ray spectrum to be concave, and rather flat at the high-energy end. The lack of such a signature in the Galactic cosmic ray spectrum has been worrisome. It appears that connecting the source spectrum to the Galactic cosmic ray spectrum as detected on Earth requires more than just taking into account energy-dependent scattering rates during propagation to our planet. In light of these results it is also unsurprising that the cumulative Galactic cosmic ray spectrum can only be steeper than $E^{-2}$. If most Galactic cosmic rays are a result of CRs escaping ahead of the SNR shock, the dynamics in itself need to be well understood, and especially the connection with the escape energy.

Our result shows that a difference in the slope can be expected depending on whether you observe cosmic rays at the source or in the far upstream environment. In order to interpret gamma-ray data it is of crucial importance to understand the exact source of the emission. For example, gamma rays coming from electrons at the shock will represent a signature of the source population, whereas gamma rays coming from escaping cosmic rays can display a different spectrum.

 \section*{Acknowledgements}
The research leading to these results has received funding
from grant number ST/H001948/1
made by the UK Science Technology and Facilities Council and from the European Research Council under the European
Community's Seventh Framework Programme (FP7/2007-2013)/ERC grant agreement no. 247039.

\bibliography{../adssample}

\label{lastpage}
\end{document}